{}%disable hyper at arxiv
{}%use pdflatex at arxiv

\documentclass[12pt,a4paper]{article}

\newif\ifpublic\publicfalse
\newif\ifniklas\niklastrue
\newif\ifniklas\niklastrue

\usepackage{ifpdf}
\ifniklas\ifpdf\publictrue\else\publicfalse
\PassOptionsToPackage{hypertex}{hyperref}
\PassOptionsToPackage{draft}{graphicx}
\fi\fi

\usepackage[utf8]{inputenc}
 \usepackage[T1]{fontenc}
\usepackage{amsmath,amssymb,amsthm}
\ifpublic\else\usepackage{showkeys}\fi
\usepackage[bookmarks=true,hyperfigures=true]{hyperref}
\usepackage{graphicx}
\usepackage{dsfont}
\usepackage{xcolor}
\usepackage[nosort]{cite}
\usepackage[bulletsep]{collref}
\ifniklas

\def\showkeysrefformat#1{{\normalfont\tiny\ttfamily#1}}
\makeatletter
\def\SK@@ref#1>#2\SK@{%
 {\@inlabelfalse\leavevmode\vbox to\z@{%
 \vss\SK@refcolor\rlap{\vrule\raise .75em%
  \hbox{\showkeysrefformat{#2}}}}}}
\makeatother
\fi

\usepackage[a4paper,text={160mm,247mm},centering]{geometry}
\allowdisplaybreaks[3]

\numberwithin{equation}{section}

\usepackage[font=small,labelfont=bf,width=0.85\textwidth]{caption}

\expandafter\def\expandafter\bfseries\expandafter{\bfseries\ifmmode\else\boldmath\fi}
\expandafter\def\expandafter\mdseries\expandafter{\mdseries\ifmmode\else\unboldmath\fi}
\expandafter\def\expandafter\normalfont\expandafter{\normalfont\ifmmode\else\unboldmath\fi}

\RequirePackage{verbatim}

\makeatletter
\newwrite\bibinl@out
\newenvironment{bibtex}[1][\jobname]{%
  \immediate\openout\bibinl@out #1.bib
  \immediate\write\bibinl@out{\@percentchar generated from `\jobname' starting line \the\inputlineno^^J}%
  \def\verbatim@processline{\immediate\write\bibinl@out{\the\verbatim@line}}%
  \@bsphack\let\do\@makeother\dospecials\catcode`\^^M\active\verbatim@start
}%
{\immediate\closeout\bibinl@out\@esphack}
\makeatother

\newcommand{\rom}[1]{\textup{\uppercase\expandafter{\romannumeral#1}}}

\usepackage{fixmath}

\newcommand{\sfrac}[2]{{\textstyle\frac{#1}{#2}}}
\newcommand{\half}{\sfrac{1}{2}}

\newcommand{\lH}{\mathcal{H}}
\newcommand{\lQ}{\mathcal{Q}}
\newcommand{\lB}{\mathcal{B}}

\newcommand{\bH}{\mathbb{H}}
\newcommand{\bQ}{\mathbb{Q}}

\RequirePackage{amsmath}
\makeatletter
\newcommand{\brk@ord}{\bBigg@{0}}
\newcommand{\brk@ordl}{\mathopen\brk@ord}
\newcommand{\brk@ordr}{\mathclose\brk@ord}
\newcommand{\brk@ordm}{\mathrel\brk@ord}
\newcommand{\brk@var}{\brk@ord}
\newcommand{\brk@varl}{\left}
\newcommand{\brk@varr}{\right}
\newcommand{\brk@varm}{\mathrel\brk@var}
\newcommand{\brk@altname}[3]{\expandafter\def\csname#2\expandafter\@gobble\string#1\endcsname{#1[#3]}}
\newcommand{\brk@usearg}[3]{%
  \def\brk@star{*}\def\brk@blank{}\def\brk@arg{#1}%
  \ifx\brk@arg\brk@blank\def\brk@arg{brk@ord}\fi%
  \ifx\brk@arg\brk@star\def\brk@arg{brk@var}\fi%
  \csname\brk@arg #2\endcsname#3}

\newcommand{\DeclareMathBrackets}[3]{
  \newcommand{#1}[2][]{\brk@usearg{##1}{l}{#2}##2\brk@usearg{##1}{r}{#3}}
  \brk@altname{#1}{big}{big}\brk@altname{#1}{lr}{*}}
\newcommand{\DeclareMathBiBrackets}[4]{
  \newcommand{#1}[3][]{\brk@usearg{##1}{l}{#2}##2#3##3\brk@usearg{##1}{r}{#4}}
  \brk@altname{#1}{big}{big}\brk@altname{#1}{lr}{*}}
\newcommand{\DeclareMathBiMBracketsStar}[4]{
  \newcommand{#1}[3][]{\brk@usearg{##1}{l}{#2}##2\brk@usearg{##1}{m}{#3}##3\brk@usearg{##1}{r}{#4}}
  \brk@altname{#1}{bi}{big}}
\newcommand{\DeclareMathBiBracketsStar}[4]{
  \newcommand{#1}[3][]{\brk@usearg{##1}{l}{#2}##2\brk@usearg{##1}{}{#3}##3\brk@usearg{##1}{r}{#4}}
  \brk@altname{#1}{big}{big}}

\makeatother

\DeclareMathBrackets{\brk}{(}{)}
\DeclareMathBrackets{\sbrk}{[}{]}
\DeclareMathBrackets{\set}{\{}{\}}
\DeclareMathBrackets{\abs}{|}{|}
\DeclareMathBrackets{\eval}{.}{|}
\DeclareMathBrackets{\spn}{\langle}{\rangle}
\DeclareMathBiBrackets{\comm}{[}{,}{]}
\DeclareMathBiBrackets{\acomm}{\{}{,}{\}}
\DeclareMathBiBrackets{\gcomm}{[}{,}{\}}

\DeclareMathOperator{\tr}{tr}

\def\[{\begin{equation}}
\def\]{\end{equation}}

\providecommand{\href}[2]{#2}

\makeatletter
\def\mr@ignsp#1 {\ifx\:#1\@empty\else #1\expandafter\mr@ignsp\fi}%
\newcommand{\multiref}[1]{\begingroup%\let\protect\string%
\xdef\mr@no@sparg{\expandafter\mr@ignsp#1 \: }%
\def\mr@comma{}%
\@for\mr@refs:=\mr@no@sparg\do{\mr@comma\def\mr@comma{,}\ref{\mr@refs}}%
\endgroup}
\renewcommand{\eqref}[1]{(\multiref{#1})}
\makeatother

\makeatletter
\newcommand{\namedref}[2]{\hyperref[#2]{#1~\ref*{#2}}}
\newcommand{\secref}{\@ifstar{\namedref{Section}}{\namedref{Sec.}}}
\newcommand{\appref}{\@ifstar{\namedref{Appendix}}{\namedref{App.}}}
\newcommand{\tabref}{\@ifstar{\namedref{Table}}{\namedref{Tab.}}}
\newcommand{\figref}{\@ifstar{\namedref{Figure}}{\namedref{Fig.}}}
\makeatother

\providecommand{\hypersetup}[1]{}

\hypersetup{plainpages=false}
\hypersetup{pdfpagemode=UseNone}
\hypersetup{bookmarksnumbered=true}
\hypersetup{pdfstartview=FitH}
\hypersetup{colorlinks=false}
\hypersetup{citebordercolor={.5 1 .5}}
\hypersetup{urlbordercolor={.5 1 1}}
\hypersetup{linkbordercolor={1 .7 .7}}
\makeatletter
\let\@keywords\@empty
\let\@subject\@empty
\providecommand{\keywords}[1]{\gdef\@keywords{#1}}
\providecommand{\subject}[1]{\gdef\@subject{#1}}
\def\thetitle{\@title}
\def\theauthor{\@author}
\def\thesubject{\@subject}
\def\thedate{\@date}
\def\thekeywords{\@keywords}
\makeatother
\AtBeginDocument{
\hypersetup{pdftitle={\thetitle}}%
\hypersetup{pdfauthor={\theauthor}}%
\hypersetup{pdfsubject={\thesubject}}%
\hypersetup{pdfkeywords={\thekeywords}}%
}

\title{Classifying integrable spin-1/2 chains with nearest neighbour interactions}
\author{ Marius de Leeuw, Anton Pribytok and Paul Ryan}

\begin{document}

\pdfbookmark[1]{Title Page}{title}
\thispagestyle{empty}

\begingroup\raggedleft\footnotesize\ttfamily
TCDMATH 19-05
\par\endgroup

\vspace*{2cm}
\begin{center}%
\begingroup\Large\bfseries\thetitle\par\endgroup
\vspace{1cm}

\begingroup\scshape\theauthor\par\endgroup
\vspace{5mm}%

\begingroup\itshape
School of Mathematics
\& Hamilton Mathematics Institute\\
Trinity College Dublin\\
Dublin, Ireland
\par\endgroup
\vspace{5mm}

\begingroup\ttfamily
mdeleeuw@maths.tcd.ie,
apribytok@maths.tcd.ie,
pryan@maths.tcd.ie
\par\endgroup

\vfill

\textbf{Abstract}\vspace{5mm}

\begin{minipage}{12.7cm}
We classify all fundamental integrable spin chains with two-dimensional local Hilbert space which have regular $R$-matrices of difference form. 
This means that the $R$-matrix underlying the integrable structures is of the form $R(u,v)=R(u-v)$ and reduces to the permutation
operator at some particular point. We find a total of 14 independent solutions, 8 of which correspond to well-known eight or
lower vertex models. The remaining 6 models appear to be new and some have peculiar properties such as not being diagonalizable
or being nilpotent. Furthermore, for even $R$-matrices, we find a bijection between solutions of the Yang-Baxter equation and the graded Yang-Baxter equation
which extends our results to the graded two-dimensional case.
\end{minipage}

\vspace*{4cm}

\end{center}

\newpage
\iffalse 
\section*{Conventions}
\textit{this is for us to be consistent}
\begin{itemize}
\item Densities are cursive $\mathcal{Q}$
\item Operators are bold $\mathbb{Q} = \sum \mathcal{Q}$
\item Eigenvalues are normal $\mathbb{Q}|u\rangle = Q |u\rangle$
\item $R$-matrix as $R$
\item Coefficients in $\lH$? we do $a,b,c,d$ or $a_1,a_2,a_3$?
\end{itemize}
\fi
%%%%%%%%%%%%%%%%%%%%%%%%%%%%%%%%%%%%%%%%%%%%%%%%%%%%%%%%%%%%%%%%%%%%%%%%%%%%%%%%
%%%%%%%%%%%%%%%%%%%%%%%%%%%%%%%%%%%%%%%%%%%%%%%%%%%%%%%%%%%%%%%%%%%%%%%%%%%%%%%%
\section{Introduction}
\label{sec:intro} 
Quantum integrable spin chains are characterised by the existence of a family of mutually commuting conserved operators 
$\bQ_2$, $\bQ_3$, $\bQ_4$, $\dots$ where the charge $\bQ_r$ acts on $r$ neighbouring sites. For an integrable system in which the interaction range of the Hamiltonian $\bH$ is two sites we naturally take $\bQ_2$ to be $\bH$ and we say that the interactions are nearest-neighbour (NN). In this case the Hamiltonian takes the form
\begin{equation}
\bH=\sum_{n=1}^L \lH_{n,n+1},
\end{equation}
where $\lH_{L,L+1}:=\lH_{L,1}$ for a length $L$ spin chain with periodic boundary conditions. The fundamental object which usually underlies the integrable structure in the theory of such models is the so-called $R$-matrix \cite{Takhtajan:1979iv,Sklyanin:1987bi}. This is an invertible operator
\begin{align}
&R_{ab}(u,v)\in {\rm End}(\mathbb{V}_a\otimes \mathbb{V}_b),&& \mathbb{V}_a\simeq \mathbb{V}_b,
\end{align}
which satisfies the Yang-Baxter equation (YBE)
\begin{equation}
R_{ab}(u,v)R_{ac}(u,w)R_{bc}(v,w)=R_{bc}(v,w)R_{ac}(u,w)R_{ab}(v,w),
\end{equation}
on $\mathbb{V}_a\otimes \mathbb{V}_b \otimes \mathbb{V}_c$. For NN interactions the $R$-matrix satisfies the \textit{regularity} condition
\begin{equation}
R_{ab}(u,u)=P_{ab},
\end{equation}
and the two-site Hamiltonian density $\lH_{12}$ can be obtained from the $R$-matrix as a logarthmic derivate
\begin{equation}
\lH_{12}=R_{12}(u,v)^{-1}\frac{dR_{12}(u,v)}{du}\Big|_{v=u}=P_{12}\frac{dR_{12}(u,v)}{du}\Big|_{v=u}.
\end{equation}
A particularly interesting class of integrable models are those where the $R$-matrix is of difference form, that is 
\begin{equation}
R_{ab}(u,v)=R_{ab}(u-v).
\end{equation}
For such models it is well-known that the tower of conserved charges $\bQ_r$, $r=2,3,\dots$ can be recursively generated by means of the so-called Boost operator $\lB[\bQ_2]$ defined by \cite{Tetelman,Loebbert:2016cdm}
\begin{equation}\label{eq:boost}
\lB[\bQ_2]:=\sum_{n=-\infty}^\infty n \lH_{n,n+1}.
\end{equation}
Defined in this way, the boost operator is only defined for infinite length chains, but its commutators with the conserved charges gives operators of finite interaction range which reduce consistently to spin chains of finite length. The boost operator can be extended to a more general setting \cite{2001PhRvL..86.5096L}.

An important and related question that we address in this paper is the classification of integrable spin chains. In general, what this means depends on one's definition of quantum integrability \cite{Caux}, but we will restrict to spin chains that have an underlying regular $R$-matrix of difference form, and we will solve the Yang-Baxter equation by making use of the boost operator. In particular we find all Hamiltonians that generate a tower of conserved charges, which is along the lines of the integrability criterion put forward in \cite{Grabowski}. For our models, we actually confirm the hypothesis from \cite{Grabowski} that having a charge $\bQ_3$ of range 3 which commutes with $\bQ_2$ is a necessary and sufficient condition for integrability.

Finding and classifying solutions of the Yang-Baxter equation is a complicated problem since it corresponds to solving a coupled set of cubic functional equations on the coefficients of the $R$-matrix. Nevertheless, certain solutions of the Yang-Baxter equation with $\mathbb{Z}_2$ symmetry were classified in \cite{Kulish1982}. Moreover, solutions of \{6,7,8\}-vertex type were classified in \cite{Sogo,Khachatryan:2012wy}. Recently all solutions of difference form for \{4,5,6,7,8\}-vertex type models have been obtained in \cite{Vieira}. Finding solutions is more tractable if one considers constant solutions. Indeed,  the  $4\times 4$ solutions of constant YBE were found in \cite{Hietarinta:1992ix}. The subset of unitary solutions \cite{2002quant.ph.11050D} and the extension to $n\times n$ solutions \cite{2018arXiv180608400P} were classified as well. %In addition to known \textit{NN}-construction, corresponding counterpart of \textit{LR} $ \mathfrak{gl} $(N) spin chains and their quantum symmetries were found \cite{beisert3001yangian} and related question of $ \mathfrak{gl} $(N) invariant $ \mathcal{R} $ matrices and master T-operator eigenvalue KP hierarchy \cite{tsuboi2015supersymmetric} (quantum-classical mappings).

 In this paper, we make use of the observation that any regular solution of the Yang-Baxter equation gives rise to an integrable spin chain with a Hamiltonian with nearest-neighbour interactions. Moreover, the tower of operators is generated by the boost operator, which, in turn, is fixed in terms of the Hamiltonian. This means that rather than solving the Yang-Baxter equations we will instead find all Hamiltonians that generate an integrable spin chain in this way. The advantage is that the Hamiltonians are constant. In this way, we have reduced the problem of finding solutions to the Yang-Baxter equation to solving a system of coupled polynomial equations rather than functional equations. For each given Hamiltonian we then subsequently derive the corresponding $R$-matrix.

Finally, we extend our analysis to the two-dimensional graded vector space $\mathbb{C}^{1|1}$. We do this by specifying a bijection between graded and non-graded solutions of the Yang-Baxter equation. In this way all our results extend to the graded case as well.

%%%%%%%%%%%%%%%%%%%%%%%%%%%%%%%%%%%%%%%%%%%%%%%%%%%%%%%%%%%%%%%%%%%%%%%%%%%%%%%%
%%%%%%%%%%%%%%%%%%%%%%%%%%%%%%%%%%%%%%%%%%%%%%%%%%%%%%%%%%%%%%%%%%%%%%%%%%%%%%%%
\section{Setting}

\paragraph{Notation and setting} 
We will be interested in classifying integrable spin chains that have an underlying $R$-matrix. In our case, the $R$-matrix is an operator $R:\mathbb{V}\otimes \mathbb{V} \rightarrow \mathbb{V}\otimes \mathbb{V} $, where $\mathbb{V} = \mathbb{C}^2$. This means that our $R$-matrix $R(u)$ is invertible and satisfies the Yang-Baxter equation
\begin{align}\label{eq:YBE}
R_{12}(u-v) R_{13}(u) R_{23}(v) =R_{23}(v) R_{13}(u)  R_{12}(u-v) .
\end{align}
Moreover, we also want $R$ to satisfy the condition that $R(0)=P$ and to be analytic in a neighbourhood of $0$. These conditions ensures that $R$ gives rise to a spin chain with a Hamiltonian with nearest-neighbour interactions.

Given such an $R$-matrix, one can construct a corresponding spin chain of length $L$. The total Hilbert space\footnote{Following standard physics terminology - we have not equipped our vector spaces with any inner product.} is then simply $\mathbb{V}^{\otimes L}$ and  the commuting conserved charges underlying the integrable structure of the spin chain are generated by the $R$-matrix via the transfer matrix
\begin{align}
t(u) = \tr_0 R_{0L}(u)\ldots R_{01}(u).
\end{align}
The trace in $t$ ensures that we are dealing with a periodic, closed spin chain, \textit{i.e.} we identify $L+1 \equiv1$. 

The Yang-Baxter equation implies that $[t(u),t(v)] = 0$ and, as a consequence, expanding the logarithm of the transfer matrix $t(u)$ around $u=0$ gives rise to a family of commuting operators $\{ \bQ_n\}_{n=1,2,\ldots}$ of increasing interaction range. The first operator corresponds to the momentum operator since $t(0)$ is simply the shift operator
\begin{align}
&t(0) = P_{LL-1}\ldots  P_{21} = e^{iP} &&\rightarrow && \bQ_1 \sim P.
\end{align}
Similarly, the next operator is usually identified as the Hamiltonian and given by the logarithmic derivative of the $R$-matrix
\begin{align}
\bQ_2 := \sum_{n=1}^L R_{n,n+1}^{-1}(0) \frac{d}{du}R_{n,n+1} \equiv \sum_n \lQ_{n,n+1},
\end{align}
where we introduced the Hamiltonian density $\lQ_{ij}$. From the definition, it is clear that the Hamiltonian only has nearest-neighbour interactions. The third charge $\bQ_3$ takes the form
\begin{align}
&\bQ_3 := \sum_n \lQ_{n,n+1,n+2},
&&  \lQ_{n,n+1,n+2} = [\lQ_{n,n+1},\lQ_{n+1,n+2}].
\end{align}
This operator can be written in terms of densities that have interaction range 3. The explicit expression for all higher operators becomes more cumbersome, but they can be elegantly described in terms of the so-called boost operator $\lB$  \cite{Tetelman,Loebbert:2016cdm}.

\paragraph{Boost operator}

The boosted Hamiltonian, $\lB[\bH]$, or $\lB[\bQ_2]$, is defined in the following way 
\begin{equation}
\lB[\bQ_2]:=\sum_{a=-\infty}^\infty a\lQ_{a,a+1},
\end{equation}
and is only defined on open spin chains of infinite length. In particular, one needs to fix an origin, \textit{i.e.} a site $0$.

The power of the boost operator comes from the fact that it can be used to recursively write all conserved charges $\bQ_r$ in the following way 
\begin{equation}
\bQ_{r+1} = [\lB[\bQ_2],\bQ_r].
\end{equation}
Hence all conserved charges can be generated just from the knowledge of the Hamiltonian $\bH$. It is easy to see that $[\lB[\bQ_2],\bQ_r]$ is a local operator of length $r+1$ and as a consequence provides a well-defined operator on any finite, periodic spin chain even though the boost operator is not defined on such models. 

%%%%%%%%%%%%%%%%%%%%%%%%%%%%%%%%%%%%%%%%%%%%%%%%%%%%%%%%%%%%%%%%%%%%%%%%%%%%%%%%
%%%%%%%%%%%%%%%%%%%%%%%%%%%%%%%%%%%%%%%%%%%%%%%%%%%%%%%%%%%%%%%%%%%%%%%%%%%%%%%%
\section{Identifications and families of solutions}\label{sec:Transf}

The total number of integrable Hamiltonians is very large. However, most Hamiltonians are part of families of solutions under certain identifications. It is easy to see that given a solution of the Yang-Baxter equation, we can generate new solutions by certain discrete and continuous transformations. Under these transformations we only need to specify a generator and then the other members of the family are easily generated by the transformations that we discuss in this section. 

\paragraph{Reduction}
A very basic way of identifying solutions is by a redefinition of the constants in the Hamiltonian or by setting some coefficients to 0. For example, the Heisenberg XXX spin chain is a reduction of the XXZ model which in turn is a special case of the XYZ spin chain. Hence, rather than listing all three models separately, we will only give the XYZ spin chain and consider the other models as special cases of this. We are also always free to choose an appropriate normalization of our Hamiltonian and add a term proportional to the identity operator, \textit{i.e.} $\mathfrak{c} \mathds{1} $.

\paragraph{Local basis transformations}
Given an $R$-matrix and the corresponding integrable spin chain, we can find a whole class of equivalent spin chains by applying a local basis transformation. If $V\in \mathrm{End}(\mathbb{V})$ is an invertible transformation then it is easy to see that $R^{(V)}(u) = (V\otimes V) R(u)  (V^{-1}\otimes V^{-1}) $ also defines an integrable spin chain, with commuting operators $\{\mathcal{Q}^{(V)}_i\}$. The map $V$ simply corresponds to a basis transformation on each site of the spin chain. As such it factorizes and we can write
\begin{align}
\mathcal{Q}^{(V)}_i = \Big(\bigotimes_L V\Big) \mathcal{Q}_i\Big(\bigotimes_L V^{-1}\Big).
\end{align}
On the level of the densities, this simply corresponds to 
\begin{align}
Q^{(V)}_{i_1,\ldots,i_r} = \Big(\bigotimes_r V\Big) Q_{i_1,\ldots,i_r}\Big(\bigotimes_r V^{-1}\Big).
\end{align}
The map $V$ is given by a $2\times2$ matrix with unit determinant. Thus, given an integrable model we can straightofrwardly generate all equivalent models under local basis transformations. Conversely, we will be looking for all integrable models and to this end we can use the degrees of freedom of a basis transformation to make sure that our Hamiltonian density is always such that certain components  vanish.

\paragraph{Discrete transformations}
It is easy to see that if $R(u)$ is a solution of the YBE then so are $PR(u)P$ and $R(u)^T$ (and it clearly follows immediately that so is $PR(u)^TP$). This means that transposition and permutation are further discrete transformations that map an integrable Hamiltonian to a different integrable Hamiltonian. 
We summarise the relation between $R$-matricies and Hamiltonians below 
\begin{align}
& R(u) &&\leftrightarrow&& \lH \\
& PR(u)P &&\leftrightarrow&& P\lH P \\
& R(u)^T &&\leftrightarrow&& P\lH^TP \\
& PR(u)^TP &&\leftrightarrow&& \lH^T
\end{align}
We emphasise that the Hamiltonian associated to $R(u)^T$ is $P\lH^TP$ and \textit{not} $\lH^T$. 

\paragraph{Equivalence classes} 
Obviously we can group Hamiltonians into equivalence classes related via the above transformations. Hence, we will not list all integrable Hamiltonians, but only a representative from each equivalence class. 

\section{Approach}

Our aim is to find all possible integrable spin chains of the type discussed above. We approach this problem from two sides. We first make a general ansatz for our Hamiltonian density and use local basis transformations to bring it into a suitable form. Assuming that there is an underlying $R$-matrix, we then use the boost operator to derive the higher operators up to $r=6$ and we demand that all commutators between those operators vanish. This will lead to a set of polynomial equations for the components of the Hamiltonian density which can be solved explicitly. 

Of course, this does not prove integrability, but for the solutions obtained in this way, we can solve the Yang-Baxter equation perturbatively. More precisely, we assume that we can expand the $R$-matrix as
\begin{align}
R = P + P\lH u + \sum_{n\geq2}R^{(n)} u^n.
\end{align}
If we plug this into the Yang-Baxter equation \eqref{eq:YBE}, we can solve for all coefficients in this power series perturbatively. For instance, we find
\begin{align}
R = P + P\lH u + P\lH^2 u^2 + \ldots.
\end{align}
The other coefficients are model dependent. Interestingly, what we notice by attempting this procedure for the XYZ model is that all of the coefficients $R^{(n)}$, $n\geq 2$ can be expressed in the following form 
\begin{equation}
R^{(n)}=P r^{(n)}(\lH),
\end{equation}
where $r^{(n)}(\lH)$ is a polynomial of degree $n$ in $\lH$. At this point we can then exploit the fact that $\lH$ is a $4\times 4$ matrix and use the Cayley-Hamilton theorem to express all higher powers of $\lH$ in terms of the identity matrix, $\lH$, $\lH^2$, and $\lH^3$, and so our ansatz can be rewritten\footnote{Note that we do not claim that \textit{all} $R$-matricies can be written in this form, but this approach turns out to be perfectly sufficient for finding the $R$-matricies of all of the new models we consider.} as 
\begin{equation}
R_{12}(u)=P_{12}\left(f_0(u)1_{12} +u f_1(u) \lH_{12}+u^2 f_2(u)\lH^2_{12} +u^3 f_3(u)\lH^3_{12}\right),
\end{equation}
which is in most cases far more convenient - we have reduced the problem to finding $4$ unknown functions of $u$, instead of an infinite number of functions of matricies $R^{(n)}$. Furthermore one of those functions can be set to $1$ by use of the gauge symmetry $R(u)\rightarrow f(u)R(u)$ of the $R$-matrix, where $f(u)$ is some analytic function. For example, we can set $f_0(u)=1$ or $f_1(u)=1$. The form of the above ansatz has also been guided by dimension analysis - the $R$-matrix should be dimensionless, but the Hamiltonian has units of energy, and the spectral parameter $u$ has units of $1/$energy. Furthermore, in order to be consistent with our general requirements the functions $f_j(u)$ should satisfy some conditions. In particular,
\begin{align}
&f_0(0)=1,
&& f'_0(0)=0,
&& f_1(0)=1.
\end{align}
Following this procedure we are able to find a corresponding $R$-matrix for \textit{each} of the new integrable Hamiltonians that we found by explicitly solving the set of equations coming from the vanishing commutators. In this way, we have proven integrability of all models that we obtained. Moreover, since any solution of the Yang-Baxter equation automatically gives rise to an integrable Hamiltonian, we have also classified regular analytic solutions of difference form of the Yang-Baxter equation. Indeed, by solving the Yang-Baxter equation perturbatively, we found that each of our Hamiltonians corresponds to a unique $R$-matrix (up to normalization).

%%%%%%%%%%%%%%%%%%%%%%%%%%%%%%%%%%%%%%%%%%%%%%%%%%%%%%%%%%%%%%%%%%%%%%%%%%%%%%%%
%%%%%%%%%%%%%%%%%%%%%%%%%%%%%%%%%%%%%%%%%%%%%%%%%%%%%%%%%%%%%%%%%%%%%%%%%%%%%%%%
\section{Computing commutators}

In this section we will spell out some of the computational details that were used to find all integrable spin chain Hamiltonians with two-dimensional local Hilbert space.

\paragraph{Ansatz}

Consider the (extended) Pauli matrices $\sigma^a$, where $b=0,\pm,3$ and $\sigma^0 = 1$, $ \sigma^{\pm} = (\sigma^{x} \pm i \sigma^{y})/2 $. These matrices form a basis for $2\times2$ matrices and hence we can write our Hamiltonian density as
\begin{align}\label{eq:AnsatzH}
\lQ_{ij} = A_{a b} \hspace{1mm} \sigma^{a} \otimes \sigma^{b}.
\end{align}
Note that we can always choose an appropriate normalization of our Hamiltonian without spoiling integrability.

Moreover, we can use equivalence under local basis transformations to set some coefficients to zero. We will attempt to set $A_{--} = A_{++} = 0$ as this simplifies our system of equations. Consider a general local basis transformation 
\begin{align}
&V = \begin{pmatrix} \alpha & \beta \\ \gamma & \delta \end{pmatrix},
&& \alpha \delta - \beta\gamma=1.
\end{align}
If we transform \eqref{eq:AnsatzH} with this basis transformation $Q^\prime = V\otimes V \cdot Q \cdot V^{-1}\otimes V^{-1}$ and look at the component in front of $\sigma^\pm \otimes \sigma^\pm$, we find
\begin{align}\label{eq:BasisGauge}
A^\prime_{++} &= \alpha^4 A_{++} \!+ \beta^4 A_{--}\! -\alpha^3 \beta (A_{+z} + A_{z+}) + \alpha \beta ^3 (A_{-z}+A_{z-}) -\alpha^2 \beta ^2 (A_{+-}+A_{-+}-A_{zz})\\
A^\prime_{--} &= \gamma^4 A_{++} \!+ \delta^4 A_{--} \!-\gamma ^3 \delta (A_{+z} + A_{z+}) + \gamma \delta ^3 (A_{-z}+A_{z-}) -\gamma ^2 \delta ^2 (A_{+-}+A_{-+}-A_{zz})
\end{align}
We see that there are three cases to consider. In the first case we take $A_{--} = A_{++} = 0$ and our Hamiltonian is already of the correct form. 

In the second case, either $A_{--}$ or $A_{++}$ is non-zero. It is then easy to see from \eqref{eq:BasisGauge} that such a matrix can be mapped to a matrix with $A^\prime_{++} = A^\prime_{--} = 0$ unless 
\begin{align}
&A_{\pm z}  = - A_{z\pm} ,
&&A_{zz}  =  A_{+-}+ A_{-+} .
\end{align}
In the final case, we take both $A_{--}$ and $A_{++}$ non-zero. But this case can always be mapped to a matrix of the second type. Hence, without loss of generality, we can restrict our Hamiltonian density to be of two types
\begin{description}
\item[Type I] Hamiltonian density has $A_{++} = A_{--} = 0$
\item[Type II] Hamiltonian density has $A_{++} =1,  A_{--} = 0$ together with $A_{\pm z}  = - A_{z\pm}$ and $A_{zz}  =  A_{+-}+ A_{-+}$
\end{description}

\paragraph{Densities}
We want to impose the vanishing of the commutators $[\bQ_r,\bQ_s]$. These operators act on the entire spin chain, but since each of these operators can be written in terms of densities of range $r,s$ respectively, we can reduce vanishing of the commutator to a condition on the corresponding densities. The computation on the level of densities reduces the problem to computing matrices in low dimensions independent of the length of the spin chain.

For concreteness, we will work out $[\bQ_2,\bQ_3]$ as the arguments straightforwardly generalize to more general commutators. The operator $\bQ_3$ has interaction range three and hence can be expressed as $\bQ_3= \sum_i A_{abd}\sigma^a_i\sigma^b_{i+1}\sigma^c_{i+2}$ for some coefficients $A_{abc}$. Thus, we can write
\begin{align}
[\bQ_2,\bQ_3] &= \sum_{n,m=1}^L A_{ab} A_{cde} [\sigma^a_n\sigma^b_{n+1} , \sigma^c_m\sigma^d_{m+1} \sigma^e_{m+2} ]\nonumber\\
&= A_{ab} A_{cde}  \sum_{n,m=1}^L (
\delta_{n+1,m}+\delta_{n,m}+\delta_{n-1,m}+\delta_{n-2,m})
[\sigma^a_n\sigma^b_{n+1} , \sigma^c_m\sigma^d_{m+1} \sigma^e_{m+2} ] \nonumber\\ 
& = B_{abcd} \sum_{n} \sigma^a_n\sigma^b_{n+1}\sigma^c_{n+2}\sigma^d_{n+3},
\end{align} 
where the $B_{abcd}$ are some combination of $A_{ab} A_{cde}$ and the structure constants of the Pauli algebra. Since $\sigma^0 = 1$, we need to consider different components of $B$ separately. For example, consider the component $B_{0bcd}$. This is actually an operator of range 3 and hence it combines with $B_{abc0}$ when the summation index is shifted by 1. Because of this, we see that we need to identify the components that have 0s on the outside indices, \textit{i.e.} $B_{0bcd} \equiv B_{bcd0}$. Doing this, we see that the commutator $[\bQ_2,\bQ_3] $ vanishes if and only if
\begin{align}
&B_{abcd} = B_{0000} = 0, \\
&B_{abc0} + B_{0abc} = 0,\\
&B_{ab00} + B_{0ab0}+ B_{00ab} = 0,\\
&B_{a000} + B_{0a00}+ B_{00a0}+ B_{000a} = 0.
\end{align}
for all $a,b,c,d\neq0$. Since the coefficients $A_{abc}$ can be expressed in terms of $A_{ab}$ via the boost operator, this means that this gives us 121 cubic equations in $A_{ab}$ that we need to solve. In general, the commutator $[\bQ_r,\bQ_s]$ will give us a set of $\half(3^{r+s-1}-1)$ polynomial equations of degree $r+s-2$. Most of these equations, such as $B_{0000}=0$ will be trivially satisfied. The vanishing of $[\bQ_2,\bQ_3]$ is equivalent to the so-called Reshetikin condition which provides a necessary condition for integrability \cite{10.1007/3-540-11190-5_8}.

\paragraph{Boost}
We would like to comment on a subtle point regarding the boost operator. Consider an operator density of the form $\delta A = A\otimes1-1\otimes A$. Summing such a density on a periodic chain means that all terms cancel $\sum_n\delta A=0$. Thus, we can add any operator of the form $A\otimes1-1\otimes A$ to the Hamiltonian density and leave the total conserved operator $\bQ_2$ unchanged. However, since the boost operator $\lB[\bQ]$ is only well-defined on open, infinite spin chains, an operator $\delta A$ will have an effect on the boosted charges and hence it will affect the form of the higher conserved charges generated from $\bQ_2$. 

\paragraph{Independent commutators}

In order to derive constraints on the coefficients of the Hamiltonian, we do not need to consider all commutators between the conserved charges $\bQ_r$. By the Jacobi identity, we find that most commutators of charges are actually related to each other. For instance, let us look at the commutator  $[\bQ_2,\bQ_4]$. Because $\bQ_4$ can be written as the commutator $[\lB[\bQ_2],\bQ_3 ]$, we find
\begin{align}
[\bQ_2,\bQ_4] =
[\bQ_2,[\lB[\bQ_2],\bQ_3]] = 
[\lB[\bQ_2],[\bQ_2,\bQ_3]] +[\bQ_3,[\lB[\bQ_2],\bQ_2]] = [\bQ_3,\bQ_3]=0,
\end{align}
since $[\bQ_2,\bQ_3]=0$. This argument can be applied inductively and we find that all commutators can be written in terms of commutators of the form
\begin{align}
&[\bQ_r,\bQ_{r+1}] 
&&\mathrm{or}
&& [\bQ_2,\bQ_{2r-1}] .
\end{align}
Thus, in what follows our aim will be to solve $[\bQ_2,\bQ_{3}]=[\bQ_3,\bQ_{4}]=0$.

%%%%%%%%%%%%%%%%%%%%%%%%%%%%%%%%%%%%%%%%%%%%%%%%%%%%%%%%%%%%%%%%%%%%%%%%%%%%%%%%
%%%%%%%%%%%%%%%%%%%%%%%%%%%%%%%%%%%%%%%%%%%%%%%%%%%%%%%%%%%%%%%%%%%%%%%%%%%%%%%%
\section{Solutions}

We begin by computing the commutator $[\bQ_2, \bQ_3]$ and find all solutions for which it vanish. Remarkably, it turns out, at least for the models considered in this paper, that demanding $[\bQ_2, \bQ_3] = 0$ is a sufficient condition to ensure the vanishing of $[\bQ_2,\bQ_4]$ and higher commutators do not yield new restrictions, confirming the hypothesis of \cite{Grabowski}. Hence in what follows we will only discuss the vanishing of $[\bQ_2,\bQ_3]$. We find in the order of 250 solutions, but most of them are equivalent up to the local basis transformations and the discrete transformations discussed in Section \ref{sec:Transf}.

The main problem is to choose an appropriate representation to present the results. We choose to use local basis transformations to make contact with the well-known examples from the literature, such as the XYZ model. However, care must be taken when doing this. For example, one solution we find is
\begin{align}
\lH = \begin{pmatrix}
a & b & -b & 0 \\
 0 & c-a & 2a- c & 0 \\
 0 & 2a+c & -a-c & 0 \\
 0 & 0 & 0 & a
\end{pmatrix}.
\end{align}
It is easy to see that the above Hamiltonian can generically be mapped to a solution of XXZ type
\begin{align}
\lH^\prime = \begin{pmatrix}
a & 0 & 0 & 0 \\
 0 & c-a & 2a-c & 0 \\
 0 & 2a+c & -a-c & 0 \\
 0 & 0 & 0 & a
\end{pmatrix}.
\end{align}
However, the similarity transformation that is needed for this takes the form
\begin{align}
V =  \begin{pmatrix}
 -\frac{2 \beta  c}{b} & \beta  \\
 0 & -\frac{b}{2 \beta  c}
\end{pmatrix}.
\end{align}
Clearly $V$ is singular when either $c=0$ or $b=0$. When $b=0$, our model is of XXZ type to start with, so no similarity transformation is needed. But when $c=0$ our solution can \textit{no longer} be brought into the form of an XYZ type model and we find a new model \eqref{eq:weirdfishnet}.

\subsection{Eight or less vertex models}

The first class we consider are Hamiltonians of XYZ type. These were already classified in \cite{Vieira}. These Hamiltonians take the general form
\begin{align}
\lH^{XYZ} = \begin{pmatrix}
a_1 & 0 & 0 & d_1 \\
0 & b_1 & c_1 & 0 \\
0 & c_2 & b_2 & 0 \\
d_2 & 0 & 0 & a_2
\end{pmatrix}.
\end{align}
There are eight independent generators of this type. For completeness, we will list these Hamiltonians explicitly

\paragraph{Diagonal (4 vertex)} Any diagonal Hamiltonian gives rise to an integrable system
\begin{align}
\mathcal{H}^{XYZ}_1 =
\begin{pmatrix}
a_1 & 0 & 0 & 0 \\
0 & b_1 & 0 & 0 \\
0 & 0 & b_2 & 0 \\
0 & 0 & 0 & a_2 
\end{pmatrix}.
\end{align}

\paragraph{XXZ} There are two families of XXZ type, which agrees with \cite{BFdLL2013integrable}
\begin{align}
&\mathcal{H}^{XYZ}_2 =
\begin{pmatrix}
a_1 & 0 & 0 & 0 \\
0 & b_1 & c_1 & 0 \\
0 & c_2 & b_2 & 0 \\
0 & 0 & 0 & a_1 
\end{pmatrix},
&&\mathcal{H}^{XYZ}_3 =
\begin{pmatrix}
a_1 & 0 & 0 & 0 \\
0 & b_1 & c_1 & 0 \\
0 & c_2 & b_2 & 0 \\
0 & 0 & 0 & -a_1-b_1-b_2 
\end{pmatrix}.
\end{align}
\paragraph{7--Vertex} There are two families of models which are of 7--vertex type
\begin{align}
&\mathcal{H}^{XYZ}_4 =
\begin{pmatrix}
a_1 & 0 & 0 & d_1 \\
0 & a_1+b_1 & c_1 & 0 \\
0 & -c_1 & a_1-b_1 & 0 \\
0 & 0 & 0 & a_1 
\end{pmatrix},
&&\mathcal{H}^{XYZ}_5 =
\begin{pmatrix}
a_1 & 0 & 0 & d_1 \\
0 & a_1- c_2 & c_1 & 0 \\
0 & c_2 & a_1- c_1 & 0 \\
0 & 0 & 0 & a_1-c_1-c_2 
\end{pmatrix}.
\end{align}
\paragraph{8--Vertex} Finally, there are three families of models which have all coefficients non-zero
\begin{align}
&\mathcal{H}^{XYZ}_6 =
\begin{pmatrix}
a_1 & 0 & 0 & d_1 \\
0 & b_1 & c_1 & 0 \\
0 & c_1 & b_1 & 0 \\
d_2 & 0 & 0 & a_1 
\end{pmatrix},
&\mathcal{H}^{XYZ}_7 =
\begin{pmatrix}
a_1 & 0 & 0 & d_1 \\
0 & b_1 & c_1 & 0 \\
0 & c_1 & b_1 & 0 \\
d_2 & 0 & 0 & 2b_1-a_1
\end{pmatrix}, \\
&\mathcal{H}^{XYZ}_8 =
\begin{pmatrix}
a_1 & 0 & 0 & d_1 \\
0 & a_1 & b_1 & 0 \\
0 & -b_1 & a_1 & 0 \\
d_2 & 0 & 0 & a_1 
\end{pmatrix}.
\end{align}
%
\iffalse
It appears that $\lH_5^{XYZ}$  is missing in the list of Hamiltonians presented in the appendix of \cite{Vieira}. It is integrable nevertheless and the corresponding $R$-matrix is
\begin{align}\label{eq:MissingVieira}
R = 
\Big[u^2 \big(a_1 - \frac{c_1+c_2}{2}\big)+u\Big]\!
\begin{pmatrix}
 \frac{e^{u c_1} c_1-e^{u c_2} c_2}{e^{u c_1}-e^{u c_2}} & 0 & 0 &
   d_1 \\
 0 & c_2 & \frac{e^{u c_2} \left(c_2-c_1\right)}{e^{u
   c_2}-e^{u c_1}} & 0 \\
 0 & \frac{e^{u c_1} \left(c_1-c_2\right)}{e^{u c_1}-e^{u c_2}} &
   c_1 & 0 \\
 0 & 0 & 0 & \frac{e^{u c_2} c_1-e^{u c_1} c_2}{e^{u c_1}-e^{u
   c_2}}
\end{pmatrix}.
\end{align}
\fi
All corresponding $R$-matrices are listed in \cite{Vieira}.

\subsection{Class 1}
The generator of the next class of Hamiltonians we find takes the form
\begin{align}\label{eq:Hnilp}
\lH_1 = \begin{pmatrix}
0 & a_1 & a_2 & 0 \\
0 & a_5 & 0 & a_3 \\
0 & 0 & -a_5 & a_4 \\
0 & 0 & 0 & 0
\end{pmatrix},
\end{align} 
where $a_1 a_3-a_2 a_4 =0$. Its $R$-matrix is given by
 \begin{align}
R_1(u) =
 \begin{pmatrix}
1 & \frac{a_1 (e^{a_5 u}-1)}{a_5}&  \frac{a_2(1-e^{-a_5 u}) }{a_5} &  \frac{a_1 a_3+a_2a_4}{a_5^2}(\cosh (a_5 u) -1) \\
0 & 0 & e^{-a_5 u} & \frac{a_4(1-e^{-a_5 u}) }{a_5} \\
0 & e^{a_5 u} & 0 & \frac{a_3 (e^{a_5 u}-1)}{a_5} \\
0 & 0 & 0 & 1
\end{pmatrix}. 
\end{align} 
It is easy to check that this $R$-matrix is regular, satisfies the Yang-Baxter equation as well as braided unitarity, $R_{12}(u)R_{21}(-u)\sim 1$. 

\subsection{Class 2}
The second class of integrable Hamiltonians is 
\begin{equation}
\lH_{2}=\left(
\begin{array}{cccc}
 0 & a_2 & a_3-a_2 & a_5 \\
 0 & a_1 & 0 & a_4 \\
 0 & 0 & -a_1 & a_3-a_4 \\
 0 & 0 & 0 & 0 \\
\end{array}
\right),
\end{equation}
which has the $R$-matrix 
\begin{align}
R_{2}(u)= u P\Big[\,\frac{a_1 }{\sinh (a_1 u)}+ \lH_{2} +\frac{ \tanh(\frac{a_1 u}{2})}{a_1}  \lH^2_{2}  \Big].
\end{align}
This $R$-matrix is regular, satisfies the Yang-Baxter equation as well as braided unitarity, $R_{12}(u)R_{21}(-u)\sim 1$.

\subsection{Class 3}

The third family of solutions is generated by
\begin{align}
\lH_{3} = 
\begin{pmatrix}
 -a_1 & \left(2 a_1-a_2\right) a_3 & \left(2 a_1+a_2\right) a_3 & 0 \\
 0 & a_1-a_2 & 0 & 0 \\
 0 & 0 & a_1+a_2 & 0 \\
 0 & 0 & 0 & -a_1 \\
\end{pmatrix},
\end{align}
which has the following $R$-matrix 
\begin{equation}
R_{3}(u)=\left(\begin{array}{cccc}
e^{-a_1 u} & a_3 \left(e^{(a_1-a_2)u}-e^{-a_1u}\right) & a_3 \left(e^{(a_1+a_2)u}-e^{-a_1u}\right) & 0 \\
0 & 0 & e^{(a_1+a_2)u} & 0 \\
0 & e^{(a_1-a_2)u} & 0 & 0 \\
0 & 0 & 0 & e^{-a_1 u}
\end{array} \right).
\end{equation}
This Hamiltonian can be seen as a deformation of a specific case of the four-vertex model, with deformation parameter $a_3$. When we set $a_3=0$ we obtain 
\begin{align}
\lH_{12} = 
\begin{pmatrix}
 -a_1 & 0 & 0 & 0 \\
 0 & a_1-a_2 & 0 & 0 \\
 0 & 0 & a_1+a_2 & 0 \\
 0 & 0 & 0 & -a_1 
\end{pmatrix},
\end{align}
which has an $R$-matrix which appeared in the classification of \cite{Vieira}. This $R$-matrix can be expressed in terms of powers of $\lH$ as 
\begin{equation}
R_{12}(u)=P_{12}(f_0(u)+uf_1(u)\lH+u^2f_2(u)\lH^2) ,
\end{equation}
where $f_j(u)$ are easily determined functions of $u,a_1,a_2$. What is rather remarkable is that the $R$-matrix is \textit{the same function} of $\lH$ for both $a_3=0$ and $a_3\neq 0$: $a_3$ enters the $R$-matrix only through the Hamiltonian, and does not appear in the coefficient functions $f_j(u)$.

\subsection{Class 4}
The next independent generator has a similar structure as $\lH_3$ and is
\begin{align}
\lH_4 = \begin{pmatrix}
a_1 & a_2 & a_2 & a_3 \\
0 & -a_1 & 0 & a_4 \\
0 & 0 & -a_1 & a_4 \\
0 & 0 & 0 & a_1
\end{pmatrix},
\end{align}
with $R$-matrix
\begin{equation}
R_{4}(u)=\left(
\begin{array}{cccc}
 e^{a_1 u } & \frac{a_2 \sinh (a_1 u )}{a_1 } & \frac{a_2 \sinh (a_1 u )}{a_1 } & \frac{e^{a_1 u } (a_2 a_4+ a_1 a_3 \coth (a_1 u )) \sinh ^2(a_1u )}{a_1 ^2} \\
 0 & 0 & e^{-a_1 u } & \frac{a_4 \sinh (a_1 u )}{a_1 } \\
 0 & e^{-a_1 u } & 0 & \frac{a_4 \sinh (a_1 u )}{a_1 } \\
 0 & 0 & 0 & e^{a_1 u } \\
\end{array}
\right).
\end{equation}
Braided unitarity is again satisfied.

\subsection{Class 5}
The fifth family has a different off-diagonal structure
\begin{equation}\label{eq:weirdfishnet}
\lH_{5} = \left(\begin{array}{cccc}
a_1 & a_2 & -a_2 & 0 \\
0 & -a_1 & 2a_1 & a_3 \\
0 & 2a_1 & -a_1 & -a_3 \\
0 & 0 & 0 & a_1
\end{array}\right).
\end{equation}
The corresponding $R$-matrix is again regular and unitary
\begin{align}
R_{5} = (1-a_1 u)\left(
\begin{array}{cccc}
 2 a_1 u+1 & a_2 u & -a_2 u & a_2 a_3 u^2 \\
 0 & 2 a_1 u & 1 & -a_3 u \\
 0 & 1 & 2 a_1 u & a_3 u \\
 0 & 0 & 0 & 2 a_1 u+1 \\
\end{array}
\right).
\end{align}

\subsection{Class 6}
The final integrable Hamiltonian is
\begin{align}
\lH_6 = \begin{pmatrix}
a_1 & a_2 & a_2 & 0 \\
0 & -a_1 & 2a_1 & -a_2 \\
0 & 2a_1 & -a_1 & -a_2 \\
0 & 0 & 0 & a_1
\end{pmatrix},
\end{align}
together with the unitary $R$-matrix
\begin{equation}
R_{6}(u)=(1-a_1 u)(1+2a_1 u)
\begin{pmatrix}
 1 & a_2 u & a_2 u & -a_2^2 u^2(2 a_1 u+1 ) \\
 0 & \frac{2 a_1 u}{2 a_1 u+1} & \frac{1}{2 a_1 u+1} & -a_2 u \\
 0 & \frac{1}{2 a_1 u+1} & \frac{2 a_1 u}{2 a_1 u+1} & -a_2 u \\
 0 & 0 & 0 & 1
\end{pmatrix}.
\end{equation}
This $R$-matrix satisfies braiding unitarity as well.

%%%%%%%%%%%%%%%%%%%%%%%%%%%%%%%%%%%%%%%%%%%%%%%%%%%%%%%%%%%%%%%%%%%%%%%%%%%%%%%%
%%%%%%%%%%%%%%%%%%%%%%%%%%%%%%%%%%%%%%%%%%%%%%%%%%%%%%%%%%%%%%%%%%%%%%%%%%%%%%%%
\section{Properties of the new models}

Let us briefly discuss some properties of the new classes of integrable models that we have encountered.
A feature which arises for generic choice of parameters in all of these models is non-diagonalisability of the corresponding Hamiltonians. In some cases this is more severe than in others - for example some of the Hamiltonians we find are nilpotent, i.e. they only have eigenvalue zero. A less severe case is those Hamiltonians which are non-diagonalisable but still contain different eigenvalues - in other words the conserved charges contain non-trivial Jordan blocks. While models with similar properties have been studied before, see \cite{Gainutdinov:2016pxy}, there has recently been a surge of interest in them due to their appearance in the so-called conformal fishnet theories \cite{Caetano:2016ydc,Gromov:2017cja,Ipsen:2018fmu}. Models with non-trivial Jordan structure also appear in the context of Temperley-Lieb or Hecke type integrable models \cite{2011JSMTE..04..007M}. However, it can be checked that none of our newly formed models fall in this category. 

\subsection{Class 1 and 2}

The conserved charges in models 1 and 2 are nilpotent. Nilpotency of the Hamiltonian is a feature of fishnet models as well \cite{Ipsen:2018fmu}.

\subsection{Class 3, 4, 5 and 6}
 
While generically these Hamiltonians are non-diagonalisable they are actually diagonalisable for certain values of the parameters. In particular,
\begin{itemize}
\item Class 3 is diagonalizable if $a_3=0$, in which case it reduces to a simple 4 vertex model.
\item Class 4 is diagonalizable if $a_2=a_4$ and $a_1a_3=a_2a_4$.
\item Class 5 is diagonalizable if $a_2+a_3=0$.
\item Class 6 is diagonalizable if $a_2=0$.
\end{itemize}
Remarkably, all eigenvalues seem to only depend on the parameter $a_1$. Hence the eigenvalues of Hamiltonians of Classes 3 and 4 correspond to the eigenvalues of the integrable spin chain with Hamiltonian density $\mathcal{H} = S^z\otimes S^z$. The eigenvalues for the spin chains of Classes 5 and 6 correspond to a spin chain with the Hamiltonian density
$\mathcal{H} = 1-2P$.

\section{Graded vector spaces}

In this section we will extend our results to $\mathbb{C}^{1|1}$. In order to fix notation, let us briefly recall some facts about graded (super) vector spaces. Let $\mathbb{V}$ be a super vector space of dimension $m|n$ - that is we have the decomposition 
\begin{align}
&\mathbb{V}=\mathbb{V}_0 \oplus \mathbb{V}_1,
&& {\rm dim}\mathbb{V}_0=m,
&&\ {\rm dim}\mathbb{V}_1=n.
\end{align}
For $i\in\{1,2,\dots,m+n \}$ let $p(i)$ denote the grading of $i$, \textit{i.e.} $p(i) = 0$ for $1\leq i\leq m$ and $p(i) = 1$ for $m+1\leq i\leq m+n$. Thus we work with the distinguished grading - $p(1)=0,\ p(2)=1$ for $\mathbb{C}^{1|1}$. By abuse of notation we will also denote the grading of any operator $X$ by $p(X)$. 

Let $E_{AB}$ denote the usual basis matrices of ${\rm End}(\mathbb{C}^{1|1})$. With the distinguished grading we have that 
\begin{equation}
p(E_{AB})=p(A)+p(B).
\end{equation}
Graded vector spaces can be equipped with a supertrace which acts on supermatrices as $\mathrm{str}\big(C_{AB}E_{AB}\big) := \sum_A (-1)^{p(A)} C_{AA} $. It satisfies the property
\begin{equation}
{\rm str}\left(XY \right)=(-1)^{p(X)p(Y)}{\rm str}\left(YX \right).
\end{equation}
Grading naturally extends to the notion of a graded tensor product $\otimes$ which satisfies $(a\otimes b)\cdot(c\otimes d) = (-1)^{|b||c|}ac\otimes bd$.

Now consider the triple graded tensor product $\mathbb{V}\otimes\mathbb{V}\otimes\mathbb{V} $ and suppose we have a graded $R$-matrix which satisfies the graded Yang-Baxter equation on this. Then, from the corresponding RTT relation it follows that 
\begin{equation}
T_a(u)T_b(u)=R_{ab}(u-v)^{-1}T_b(u)T_a(u)R_{ab}(u-v).
\end{equation}
In graded integrable models, the transfer matrix is defined as the supertrace of the monodromy matrix $t(u) = \mathrm{str} T(u)$. Thus, we see that if $R$ is not an even operator, then it does not follow that 
\begin{equation}
t(u)t(v)=t(v)t(u).
\end{equation}
In other words, only even solutions of the Yang-Baxter equation will generate a tower of conserved charges associated with integrability. Of course, from a physical perspective it makes little sense to consider Hamiltonians corresponding to odd operators.

Hence, while it is of course possible to have generic solutions of the Yang-Baxter equation only even solutions guarantee that the charges generated by the transfer matrix are commutative, and hence we restrict our attention to these solutions. This means that the relevant Hamiltonians will be of XYZ type. These matrices are interesting for example for the scattering of massless excitations in the AdS/CFT correspondence \cite{Fontanella:2017rvu,Bombardelli:2018jkj}.

\paragraph{Integrable systems} By making all tensor products graded, our analysis can straightforwardly be extended to the graded case, which was worked out in \cite{Kulish:1985bj,Bracken:1994hz, batchelor2008quantum}. We find the same solutions for integrable Hamiltonians, up to an extra sign in the front of the term $E_{21}\otimes E_{12}$. Remarkably, in the usual convention of matrix representation of graded operators
\begin{equation}
A=\sum A_{ijkl} E_{ij}\otimes E_{kl}(-1)^{(p(i)+p(j))p(k)}\quad \rightarrow \quad 
\left(\begin{array}{cccc}
A_{1111} & A_{1112} & A_{1211} & A_{1212} \\
A_{1121} & A_{1122} & A_{1221} & A_{1222} \\
A_{2111} & A_{2112} & A_{2211} & A_{2212} \\
A_{2121} & A_{2122} & A_{2221} & A_{2222}
\end{array}\right),
\end{equation}
this results in exactly the same generators as in the even case (after some redefinitions).

This can be seen in two different ways. First, we have derived this by direct computation by finding the integrable Hamiltonians. Second, any given solution of the Yang-Baxter equation of XYZ type for $\mathbb{C}^2$ can be mapped to a solution of the graded Yang-Baxter equation. Specifically, if 
\begin{equation}
R(u)=\left(\begin{array}{cccc}
a_1(u) & 0 & 0 & d_1 (u) \\
0 & b_1(u) & c_1(u) & 0 \\
0 & c_2(u) & b_2(u) & 0 \\
d_2(u) & 0 & 0 & a_2(u)
\end{array}\right),
\end{equation} 
is a regular solution of the Yang-Baxter equation then 
\begin{equation}
R(u)=\left(\begin{array}{cccc}
a_1(u) & 0 & 0 & \epsilon_1 d_1 (u) \\
0 & \epsilon_2 b_1(u) & c_1(u) & 0 \\
0 & c_2(u) & \epsilon_2 b_2(u) & 0 \\
-\epsilon_1 d_2(u) & 0 & 0 & -a_2(u)
\end{array}\right),
\end{equation}
is a regular solution of the graded Yang-Baxter equation, where $\epsilon_i \in \{-1,+1 \}$. It may seem like one can obtain a number of different graded solutions from a non-graded one, but this is simply due to the fact that if 
\begin{equation}
R(u)=\left(\begin{array}{cccc}
a_1(u) & 0 & 0 & d_1 (u) \\
0 & b_1(u) & c_1(u) & 0 \\
0 & c_2(u) & b_2(u) & 0 \\
d_2(u) & 0 & 0 & a_2(u)
\end{array}\right),
\end{equation} 
is a regular solution of the YBE then so is 
\begin{equation}
R(u)=\left(\begin{array}{cccc}
a_1(u) & 0 & 0 & \epsilon_1 d_1 (u) \\
0 & \epsilon_2 b_1(u) & c_1(u) & 0 \\
0 & c_2(u) & \epsilon_2 b_2(u) & 0 \\
\epsilon_1 d_2(u) & 0 & 0 & a_2(u)
\end{array}\right),
\end{equation}
and hence the map 
\begin{equation}
\left(\begin{array}{cccc}
a_1(u) & 0 & 0 & d_1 (u) \\
0 & b_1(u) & c_1(u) & 0 \\
0 & c_2(u) & b_2(u) & 0 \\
d_2(u) & 0 & 0 & a_2(u)
\end{array}\right)\quad  \rightarrow\quad  \left(\begin{array}{cccc}
a_1(u) & 0 & 0 & d_1 (u) \\
0 & b_1(u) & c_1(u) & 0 \\
0 & c_2(u) & b_2(u) & 0 \\
- d_2(u) & 0 & 0 & - a_2(u)
\end{array}\right),
\end{equation}
defines a bijection between regular solutions of the YBE and regular solutions of the graded YBE (of XYZ type).

\section{Conclusions and Discussion}

In this paper we classified all analytic difference form solutions of the Yang-Baxter equation - equivalently all nearest-neighbour spin chains with periodic boundary conditions for which the tower of conserved charges can be generated from the Hamiltonian by means of the boost operator. Our approach was based on making a general ansatz for the Hamiltonian $\mathbb{Q}_2$ and then solving the resulting polynomial equations stemming from the requirement that $[\mathbb{Q}_2,\mathbb{Q}_3]=0$. In each of these cases we were then able to find explicit $R$-matrices, ensuring the integrability of the models. As a result of this approach we found a number of solutions which have not previously appeared in the literature. 

There are a number of interesting directions one could look at for further study. Firstly, it would be interesting to study the quantum algebras associated with each of the new $R$-matrices as well as the dynamics of each of the physical models. These should be complimentary, and it would be very interesting to develop some Bethe Ansatz-like techniques to study the spectrum and eigenstates of the conserved charges. Furthermore, the non-diagonalisable Hamiltonians we found could provide a useful playground for developing techniques which may be subsequently applicable to other models with non-diagonalisable Hamiltonians, such as the conformal fishnet theories \cite{Caetano:2016ydc,Gromov:2017cja,Ipsen:2018fmu}. Another potential application would be a generalisation to open spin chains to find solutions of the boundary Yang-Baxter equation \cite{BFdLL2013integrable,Loebbert:2016cdm}. 

We have also classified the regular two-dimensional supersymmetric spin chains. Indeed, as the Hamiltonian density and $R$-matrices must be even operators it follows that all Hamiltonians must be at most of XYZ-type. For this case we were able to formulate a bijection between the graded and non-graded solutions. 

A natural extension of this work is the classification of (supersymmetric) spin chains with local spin sites of dimension $d>2$. The latter task is likely to be quite involved. The Hamiltonian density for the generic $d=3$ case can contain up to $9^2=81$ free parameters, before applying integrability-preserving transformations. Furthermore, it would also be considerably more difficult to construct the corresponding $R$-matrices. In this paper we exploited the fact that the Hamiltonian density was a $4\times 4$ matrix and used the Cayley-Hamilton theorem to write an ansatz for $R$ in terms of $1,\lH,\lH^2,\lH^3$, which turned out to be a rather efficient approach. However, if one was to apply a similar approach to higher-rank models, it would in principle be necessary to include all powers $\lH^j$, $j=0,1,\dots 8$. For models where this ansatz is not applicable, one would have to solve the YBE perturbatively in order to see if there are any simplifications in the $R$-matrix which can be exploited. For example, it may be that many of the entries can be set to zero. Once this is known, it would then remain to solve the functional relations stemming from the YBE. Nevertheless, this method will probably be applicable when one considers models with additional restrictions/symmetries such that the number of free components in the Hamiltonian will be reduced. It would be also interesting to see if the relation between graded and non-graded solutions can be generalized to spin chains of higher dimensions.

\paragraph{Acknowledgements.}

We would like to thank B. Basso, S. Frolov, F. Loebbert, T. McLoughlin, A. Retore, I. Runkel, A. Torrielli, R. Vieira and D. Volin for useful discussions. MdL was supported by SFI and the Royal Society for funding under grants UF160578 and RGF$\backslash$EA$\backslash$180167. A.P. is also supported by the grant RGF$\backslash$EA$\backslash$180167. The work of P.R. is supported in part by a Nordita Visiting PhD Fellowship. 

%%%%%%%%%%%%%%%%%%%%%%%%%%%%%%%%%%%%%%%%%%%%%%%%%%%%%%%%%%%%%%%%%%%%%%%%%%%%%%%%
%%%%%%%%%%%%%%%%%%%%%%%%%%%%%%%%%%%%%%%%%%%%%%%%%%%%%%%%%%%%%%%%%%%%%%%%%%%%%%%%

\begin{bibtex}[\jobname]

@article{Tetelman,
      author         = "Tetelman, M.G.",
      title          = "{ Lorentz group for two-dimensional integrable lattice systems.}",
      journal        = "Sov. Phys. JETP",
      volume         = "55(2)",
      year           = "1982",
      pages          = "306-310",
      doi            = "",
      reportNumber   = "2",
      SLACcitation   = ""
}

@article{Loebbert:2016cdm,
      author         = "Loebbert, Florian",
      title          = "{Lectures on Yangian Symmetry}",
      journal        = "J. Phys.",
      volume         = "A49",
      year           = "2016",
      number         = "32",
      pages          = "323002",
      doi            = "10.1088/1751-8113/49/32/323002",
      eprint         = "1606.02947",
      archivePrefix  = "arXiv",
      primaryClass   = "hep-th",
      reportNumber   = "HU-EP-16-12",
      SLACcitation   = "%%CITATION = ARXIV:1606.02947;%%"
}

@article{Vieira,
      author         = "Vieira, R. S.",
      title          = "{Solving and classifying the solutions of the Yang-Baxter
                        equation through a differential approach. Two-state
                        systems}",
      journal        = "JHEP",
      volume         = "10",
      year           = "2018",
      pages          = "110",
      doi            = "10.1007/JHEP10(2018)110",
      eprint         = "1712.02341",
      archivePrefix  = "arXiv",
      primaryClass   = "nlin.SI",
      SLACcitation   = "%%CITATION = ARXIV:1712.02341;%%"
}

@article{BFdLL2013integrable,
	title={Integrable deformations of the XXZ spin chain},
	author={Beisert, Niklas and Fi{\'e}vet, Lucas and de Leeuw, Marius and Loebbert, Florian},
	journal={Journal of Statistical Mechanics: Theory and Experiment},
	volume={2013},
	number={09},
	pages={P09028},
	year={2013},
	publisher={IOP Publishing}
}

@book{Marshakov1999SeibergWitten,
	title={Seiberg-Witten theory and integrable systems},
	author={Marshakov, Andrei},
	year={1999},
	publisher={World Scientific}
}

@article{Ipsen:2018fmu,
      author         = "Ipsen, Asger C. and Staudacher, Matthias and Zippelius,
                        Leonard",
      title          = "{The one-loop spectral problem of strongly twisted $
                        \mathcal{N} $ = 4 Super Yang-Mills theory}",
      journal        = "JHEP",
      volume         = "04",
      year           = "2019",
      pages          = "044",
      doi            = "10.1007/JHEP04(2019)044",
      eprint         = "1812.08794",
      archivePrefix  = "arXiv",
      primaryClass   = "hep-th",
      reportNumber   = "HU-Mathematik-2018-11, HU-EP-18/39",
      SLACcitation   = "%%CITATION = ARXIV:1812.08794;%%"
}

@article{Gainutdinov:2016pxy,
      author         = "Gainutdinov, Azat M. and Nepomechie, Rafael I.",
      title          = "{Algebraic Bethe ansatz for the quantum group invariant
                        open XXZ chain at roots of unity}",
      journal        = "Nucl. Phys.",
      volume         = "B909",
      year           = "2016",
      pages          = "796-839",
      doi            = "10.1016/j.nuclphysb.2016.06.007",
      eprint         = "1603.09249",
      archivePrefix  = "arXiv",
      primaryClass   = "math-ph",
      reportNumber   = "UMTG-283",
      SLACcitation   = "%%CITATION = ARXIV:1603.09249;%%"
}

@article{Caux,
	year = 2011,
	month = {feb},
	publisher = {{IOP} Publishing},
	volume = {2011},
	number = {02},
	pages = {P02023},
	author = {Jean-Sebastien Caux and Jorn Mossel},
	title = {Remarks on the notion of quantum integrability},
	journal = {Journal of Statistical Mechanics: Theory and Experiment},
}

@article{Grabowski,
	year = 1995,
	month = {sep},
	publisher = {{IOP} Publishing},
	volume = {28},
	number = {17},
	pages = {4777--4798},
	author = {M P Grabowski and P Mathieu},
	title = {Integrability test for spin chains},
	journal = {Journal of Physics A: Mathematical and General},
}

@Article{Kulish1982,
author="Kulish, P. P.
and Sklyanin, E. K.",
title="Solutions of the Yang-Baxter equation",
journal="Journal of Soviet Mathematics",
year="1982",
month="Jul",
day="01",
volume="19",
number="5",
pages="1596--1620",
}

@article{Fontanella:2017rvu,
      author         = "Fontanella, Andrea and Torrielli, Alessandro",
      title          = "{Massless $AdS_2$ scattering and Bethe ansatz}",
      journal        = "JHEP",
      volume         = "09",
      year           = "2017",
      pages          = "075",
      doi            = "10.1007/JHEP09(2017)075",
      eprint         = "1706.02634",
      archivePrefix  = "arXiv",
      primaryClass   = "hep-th",
      reportNumber   = "DMUS-MP-17-05",
      SLACcitation   = "%%CITATION = ARXIV:1706.02634;%%"
}

@Article{2002quant.ph.11050D,
author="Dye, H. A.",
title="Unitary Solutions to the Yang-Baxter Equation in Dimension Four",
journal="Quantum Information Processing",
year="2003",
month="Apr",
day="01",
volume="2",
number="1",
pages="117--152",
abstract="In this paper, we determine all unitary solutions to the Yang-Baxter equation in dimension four. Quantum computation motivates this study. This set of solutions will assist in clarifying the relationship between quantum entanglement and topological entanglement. We present a variety of facts about the Yang-Baxter equation for the reader unfamiliar with the equation.",
issn="1573-1332",
doi="10.1023/A:1025843426102",
url="https://doi.org/10.1023/A:1025843426102"
}

@article{Hietarinta:1992ix,
      author         = "Hietarinta, Jarmo",
      title          = "{Solving the two-dimensional constant quantum Yang-Baxter
                        equation}",
      journal        = "J. Math. Phys.",
      volume         = "34",
      year           = "1993",
      pages          = "1725-1756",
      doi            = "10.1063/1.530185",
      reportNumber   = "TURKU-FL-R7",
      SLACcitation   = "%%CITATION = JMAPA,34,1725;%%"
}

@ARTICLE{2018arXiv180608400P,
       author = {{Pourkia}, Arash},
        title = "{Solutions to the constant Yang-Baxter equation in all dimensions}",
      journal = {arXiv e-prints},
     keywords = {Quantum Physics},
         year = "2018",
        month = "Jun",
          eid = {arXiv:1806.08400},
        pages = {arXiv:1806.08400},
archivePrefix = {arXiv},
       eprint = {1806.08400},
 primaryClass = {quant-ph},
       adsurl = {https://ui.adsabs.harvard.edu/abs/2018arXiv180608400P},
      adsnote = {Provided by the SAO/NASA Astrophysics Data System}
}

@article{Bombardelli:2018jkj,
      author         = "Bombardelli, Diego and Stefanski, Bogdan and Torrielli,
                        Alessandro",
      title          = "{The low-energy limit of AdS$_{3}$/CFT$_{2}$ and its
                        TBA}",
      journal        = "JHEP",
      volume         = "10",
      year           = "2018",
      pages          = "177",
      doi            = "10.1007/JHEP10(2018)177",
      eprint         = "1807.07775",
      archivePrefix  = "arXiv",
      primaryClass   = "hep-th",
      reportNumber   = "DMUS-MP-18-04",
      SLACcitation   = "%%CITATION = ARXIV:1807.07775;%%"
}

@article{Gromov:2017cja,
      author         = "Gromov, Nikolay and Kazakov, Vladimir and Korchemsky,
                        Gregory and Negro, Stefano and Sizov, Grigory",
      title          = "{Integrability of Conformal Fishnet Theory}",
      journal        = "JHEP",
      volume         = "01",
      year           = "2018",
      pages          = "095",
      doi            = "10.1007/JHEP01(2018)095",
      eprint         = "1706.04167",
      archivePrefix  = "arXiv",
      primaryClass   = "hep-th",
      SLACcitation   = "%%CITATION = ARXIV:1706.04167;%%"
}

@article{Caetano:2016ydc,
      author         = "Caetano, Joao and Gurdogan, Omer and Kazakov,
                        Vladimir",
      title          = "{Chiral limit of $ \mathcal{N} $ = 4 SYM and ABJM and
                        integrable Feynman graphs}",
      journal        = "JHEP",
      volume         = "03",
      year           = "2018",
      pages          = "077",
      doi            = "10.1007/JHEP03(2018)077",
      eprint         = "1612.05895",
      archivePrefix  = "arXiv",
      primaryClass   = "hep-th",
      SLACcitation   = "%%CITATION = ARXIV:1612.05895;%%"
}

@article{tsuboi2015supersymmetric,
	title={Supersymmetric quantum spin chains and classical integrable systems},
	author={Tsuboi, Zengo and Zabrodin, Anton and Zotov, Andrei},
	journal={Journal of High Energy Physics},
	volume={2015},
	number={5},
	pages={86},
	year={2015},
	publisher={Springer}
}

@article{basso2019continuum,
	title={Continuum limit of fishnet graphs and AdS sigma model},
	author={Basso, Benjamin and Zhong, De-liang},
	journal={Journal of High Energy Physics},
	volume={2019},
	number={1},
	pages={2},
	year={2019},
	publisher={Springer}
}

@article{loebbert2012recursion,
	title={Recursion relations for long-range integrable spin chains with open boundary conditions},
	author={Loebbert, Florian},
	journal={Physical Review D},
	volume={85},
	number={8},
	pages={086008},
	year={2012},
	publisher={APS}
}

@article{Beisert:2007jv,
	author         = "Beisert, Niklas and Erkal, Denis",
	title          = "{Yangian symmetry of long-range gl(N) integrable spin
		chains}",
	journal        = "J. Stat. Mech.",
	volume         = "0803",
	year           = "2008",
	pages          = "P03001",
	doi            = "10.1088/1742-5468/2008/03/P03001",
	eprint         = "0711.4813",
	archivePrefix  = "arXiv",
	primaryClass   = "hep-th",
	reportNumber   = "AEI-2007-166, EFI-07-36, PUTP-2234",
	SLACcitation   = "%%CITATION = ARXIV:0711.4813;%%"
}
@article{Khachatryan:2012wy,
      author         = "Khachatryan, Sh. and Sedrakyan, A.",
      title          = "{On the solutions of the Yang-Baxter equations with
                        general inhomogeneous eight-vertex $R$-matrix: Relations
                        with Zamolodchikov's tetrahedral algebra}",
      journal        = "J. Statist. Phys.",
      volume         = "150",
      year           = "2013",
      pages          = "130",
      doi            = "10.1007/s10955-012-0666-8",
      eprint         = "1208.4339",
      archivePrefix  = "arXiv",
      primaryClass   = "math-ph",
      SLACcitation   = "%%CITATION = ARXIV:1208.4339;%%"
}
@article{Sogo,
      author         = "K. Sogo and M. Uchinami. Y and Akutsu and M. Wadat",
      title          = "{Classification of Exactly Solvable Two-Component Models}",
      journal        = "Progress of Theoretical Physics",
      volume         = "68",
      year           = "1982",
      pages          = "508-526",
      doi            = "10.1143/PTP.68.508"
}
@article{Takhtajan:1979iv,
      author         = "Takhtajan, L. A. and Faddeev, L. D.",
      title          = "{The Quantum method of the inverse problem and the
                        Heisenberg XYZ model}",
      journal        = "Russ. Math. Surveys",
      volume         = "34",
      year           = "1979",
      number         = "5",
      pages          = "11-68",
      note           = "[Usp. Mat. Nauk34,no.5,13(1979)]",
      SLACcitation   = "%%CITATION = RMSUA,34,11;%%"
}
@article{Sklyanin:1987bi,
      author         = "Sklyanin, E. K.",
      title          = "{Boundary conditions for integrable equations}",
      journal        = "Funct. Anal. Appl.",
      volume         = "21",
      year           = "1987",
      pages          = "164-166",
      doi            = "10.1007/BF01078038",
      note           = "[Funkt. Anal. Pril.21N2,86(1987)]",
      SLACcitation   = "%%CITATION = FAAPB,21,164;%%"
}
@ARTICLE{2001PhRvL..86.5096L,
       author = {{Links}, Jon and {Zhou}, Huan-Qiang and {McKenzie}, Ross H. and
         {Gould}, Mark D.},
        title = "{Ladder Operator for the One-Dimensional Hubbard Model}",
      journal = {Physical Review Letters},
     keywords = {Condensed Matter - Strongly Correlated Electrons},
         year = "2001",
        month = "May",
       volume = {86},
       number = {22},
        pages = {5096-5099},
          doi = {10.1103/PhysRevLett.86.5096},
archivePrefix = {arXiv},
       eprint = {cond-mat/0011368},
 primaryClass = {cond-mat.str-el},
       adsurl = {https://ui.adsabs.harvard.edu/abs/2001PhRvL..86.5096L},
      adsnote = {Provided by the SAO/NASA Astrophysics Data System}
}

@article{Kulish:1985bj,
      author         = "Kulish, P. P.",
      title          = "{Integrable graded magnets}",
      journal        = "J. Sov. Math.",
      volume         = "35",
      year           = "1986",
      pages          = "2648-2662",
      doi            = "10.1007/BF01083770",
      note           = "[Zap. Nauchn. Semin.145,140(1985)]",
      SLACcitation   = "%%CITATION = JOSMA,35,2648;%%"
}
@article{Bracken:1994hz,
      author         = "Bracken, Anthony J. and Gould, Mark D. and Zhang,
                        Yao-Zhong and Delius, Gustav W.",
      title          = "{Solutions of the quantum Yang-Baxter equation with extra
                        nonadditive parameters}",
      journal        = "J. Phys.",
      volume         = "A27",
      year           = "1994",
      pages          = "6551-6562",
      doi            = "10.1088/0305-4470/27/19/025",
      eprint         = "hep-th/9405138",
      archivePrefix  = "arXiv",
      primaryClass   = "hep-th",
      reportNumber   = "UQMATH-94-03, BI-TP-94-23",
      SLACcitation   = "%%CITATION = HEP-TH/9405138;%%"
}
@article{batchelor2008quantum,
  title={The quantum inverse scattering method with anyonic grading},
  author={Batchekor, M and Foerster, A and Guan, X-W and Links, J and Zhou, H-Q},
  journal={Journal of physics. A, Mathematical and theoretical},
  volume={41},
  number={46},
  year={2008},
  publisher={IOP}
}

@InProceedings{10.1007/3-540-11190-5_8,
author="Kulish, P. P.
and Sklyanin, E. K.",
editor="Hietarinta, J.
and Montonen, C.",
title="Quantum spectral transform method recent developments",
booktitle="Integrable Quantum Field Theories",
year="1982",
publisher="Springer Berlin Heidelberg",
address="Berlin, Heidelberg",
pages="61--119",
isbn="978-3-540-38976-7"
}

@ARTICLE{2008JMP....49b3510K,
       author = {{Kulish}, P.~P. and {Manojlovic}, N. and {Nagy}, Z.},
        title = "{Quantum symmetry algebras of spin systems related to Temperley-Lieb R-matrices}",
      journal = {Journal of Mathematical Physics},
     keywords = {Mathematics - Quantum Algebra, Nonlinear Sciences - Exactly Solvable and Integrable Systems, 81R12, 81R50, 17B37},
         year = "2008",
        month = "Feb",
       volume = {49},
       number = {2},
          eid = {023510},
        pages = {023510},
          doi = {10.1063/1.2873025},
archivePrefix = {arXiv},
       eprint = {0712.3154},
 primaryClass = {math.QA},
       adsurl = {https://ui.adsabs.harvard.edu/abs/2008JMP....49b3510K},
      adsnote = {Provided by the SAO/NASA Astrophysics Data System}
}

@ARTICLE{2011JSMTE..04..007M,
       author = {{Morin-Duchesne}, Alexi and {Saint-Aubin}, Yvan},
        title = "{The Jordan structure of two-dimensional loop models}",
      journal = {Journal of Statistical Mechanics: Theory and Experiment},
     keywords = {Mathematical Physics, Condensed Matter - Statistical Mechanics},
         year = "2011",
        month = "Apr",
       volume = {2011},
       number = {4},
        pages = {04007},
          doi = {10.1088/1742-5468/2011/04/P04007},
archivePrefix = {arXiv},
       eprint = {1101.2885},
 primaryClass = {math-ph},
       adsurl = {https://ui.adsabs.harvard.edu/abs/2011JSMTE..04..007M},
      adsnote = {Provided by the SAO/NASA Astrophysics Data System}
}

\end{bibtex}

\bibliographystyle{nb}
\bibliography{\jobname}

\end{document}